\documentclass{appolb}
\preprint{}
\usepackage{amsmath,epsfig}

\begin{document}

\title{Tensor and vector formulations of\\
resonance effective theory \thanks{Presented by K.K. at the Final Euridice
Meeting, 24-27 August 2006, Kazimierz}
\thanks{This work is supported in part by the Center for Particle Physics
(project no. LC 527) and EU RTN programme EURIDICE.} }
\author{Karol Kampf, Ji\v{r}\'{\i} Novotn\'y, Jaroslav Trnka
\address{Institute of Particle and Nuclear Physics, Faculty of Mathematics and Physics,
Charles University, V Hole\v{s}ovi\v{c}k\'ach 2, CZ - 180 00
Prague 8, Czech republic}}
\maketitle

\begin{abstract}
The main idea of the \emph{first order formalism} is demonstrated on a toy
example of spin-0 particle. The full formalism for spin-1 is applied to the
vector formfactor of the pion and its high-energy behaviour is studied.
\end{abstract}


\section{Introduction}

Recently, a considerable progress in the resonance saturation of the $O(p^6)$
low energy constants (LECs) of the chiral perturbation theory ($\chi PT$)
Lagrangian has been made. While a systematic classification of the relevant
operator basis entering the large $N_C$ $QCD$ motivated Lagrangian of the
chiral resonance theory ($\chi RT$) along with the formal resonance saturation
of the $O(p^6)$ LECs was given in \cite{CEEKPP}, a phenomenological analysis
for the symmetry breaking sector and a numerical estimate of the vector
resonance contribution was discussed in detail in \cite{moussallam}. The
importance of the vector resonances is established in the $O(p^4)$ case, where
whenever the vector resonances contribute, they dominate \cite{ecker1}. Note
however, that the field-theory formulation of this sector of $\chi RT$ is not
unique. The reason is that spin-1 resonances can be described using various
types of fields, the most common ones in this context are the vector Proca
fields and antisymmetric tensor fields (both formalisms will be discussed
briefly in Section \ref{Secpt}). Physical equivalence of various
formulation of the spin-1 sector of $\chi RT$ at the order $O(p^4)$ was studied
already in the seminal paper \cite{ecker2}, leading to the necessity to append
additional contact terms to the Proca field $\chi RT$ Lagrangian in order to
ensure both the \emph{formal} equivalence with the antisymmetric tensor
description (meaning equality of the correlators) and the \emph{physical} one
(to satisfy the high energy constraints dictated by $QCD$). At the $O(p^6)$,
the situation is more complex \cite{nase}. To obtain complete \emph{formal}
equivalence, one needs to add an infinite number of terms with increasing
chiral order (to the contrary to $O(p^4)$ case not only contact terms but also
terms with explicit resonance fields). This infinite tower of operators can be
truncated, provided we confine ourselves to the saturation of the $O(p^6)$ LECs
or provided we require equality of the $n$-point correlators up to some
$n_{\max}$ only (see \cite{nase} for details). Even in such a case, however,
both formalisms are in a sense incomplete, because generally some contributions
to the $O(p^6)$ LECs are still missing in both of them and have to be added by
hand (for the tensor formalism this was made in \cite{CEEKPP} and
\cite{moussallam} by means of the comparison with Proca field one). The
necessity of additional terms might be required by the high energy $QCD$
constraints \cite{CEEKPP}, however, the exhausting analysis is still missing
here.

In \cite{nase}, another possibility to describe spin-1 resonances has been
discussed, which gives automatically all the contributions to the $O(p^6)$
LECs known from the both formalisms mentioned above and which is in some
sense a synthesis of them. Within this formalism, the spin-1 particles are
described by a \emph{pair} of vector and antisymmetric tensor fields which
satisfy the first order equations of motion.

The aim of this paper is to demonstrate the main idea of this \emph{first
order formalism} using a toy example (see Section \ref{Secfof}) and also to
illustrate the application of the full formalism in the concrete case of the
vector formfactor of the pion (see Section \ref{Secex}).

\section{Description of vector resonances}

\label{Secpt}

Let us start with a brief review of two standard representations for the
$1^{--}$ resonances. The first one, known from textbooks, is the familiar Proca
field formalism; the (nonet) vector degree of freedom is represented by means
of Lorentz vector $V_\mu^a$ (using Gell-Mann and unit matrices we have
$V_\mu=\lambda^a V_\mu^a /\sqrt2$). The free field Lagrangian is given by
\begin{equation}
\mathcal{L}^P_0 = -\tfrac14 \langle V_{\mu\nu}V^{\mu\nu} - 2 M_V^2 V_\mu
V^\mu\rangle\,,
\end{equation}
where $V_{\mu\nu} = \partial_\mu V_\nu - \partial_\nu V_\mu$ and
$\langle\ldots\rangle$ represents a trace in the flavour space.

The second formalism uses an antisymmetric tensor $R^a_{\mu\nu}$ with
\begin{equation}
\mathcal{L}^T_0 = -\tfrac12 \langle \partial^\lambda R_{\lambda \mu}
\partial_\nu R^{\nu\mu} - \tfrac12 M_V^2 R_{\mu\nu}R^{\mu\nu} \rangle\,.
\end{equation}
As follows from the equations of motion, this choice of the Lagrangian
implies that the $R^{ij}$ degrees of freedom do not propagate. It could be
shown that these two Lagrangians are equivalent (at least at the classical
level).

The interactions with the Goldstone bosons follow from the symmetry properties
of the underlying theory ($QCD$). Resonances transform as nonet under the
$U(3)_V$ in the non-linear realization of chiral symmetry. The full Lagrangian
can be constructed using the standard chiral building blocks of this non-linear
realization together with the resonance fields. The leading order is
represented by monomials which are linear in the resonance fields and chiral
blocks are fixed by the standard ChPT counting scheme. These leading order
monomials were first studied in \cite{ecker2,prades} and \cite{ecker1} for
Proca and tensor formalism, respectively.

Integrating out the resonances one can study directly their influence in the
sector of Goldstone bosons by saturation of low energy constants of $\chi PT$.
This can also tell us what is the order of resonance nonet from the point of
view of $\chi PT$; one gets: $O(p^3)$ for $V^\mu$ and $O(p^2)$ for
$R^{\mu\nu}$. Having this powercounting one can construct further higher order
terms; for this we refer to \cite{prades, CEEKPP, moussallam} (see also
\cite{nase} and references therein).

As mentioned in Introduction, for a given order the equivalence between
$\mathcal{L}^P$ and $\mathcal{L}^T$ is lost. One can ask which formalism is
more appropriate to describe the vector resonances. Intuitively, due to the
lower counting and thus richer structure, one would vote for tensor formalism
(\emph{e.g.} in Proca formalism we have no contribution to $O(p^4)$ LECs). The
situation is, however, more complicated. The point is that, even in this case,
one is forced to fix extra contact terms (\emph{i.e.} terms without resonances) in
order to make the theory consistent. For instance (cf. \cite{CEEKPP}) at the
order $O(p^6)$, the Froissart bound applied to the $\pi^0$ Compton scattering
amplitude calculated within the tensor formalism requires to add one such
contact term. What is, however, interesting is that this term arises naturally
in Proca formalism. It would be therefore useful to formulate the resonance
theory in a way which would take advantages of both formalisms at once. This
can be done using the first order formalism with the (free) Lagrangian of the
form
\begin{equation}
\mathcal{L}^{PT}_0=\frac 14M_V^2\left\langle R_{\mu \nu }R^{\mu \nu
}\right\rangle +\frac 12M_V^2\left\langle V_\mu V^\mu \right\rangle -\frac
12M_V\left\langle V_{\mu \nu }R^{\mu \nu }\right\rangle\,,
\end{equation}
which was introduced in \cite{nase}.

\section{First order formalism}

\label{Secfof}

In order to avoid cumbersome expressions with lots of indices, let
us demonstrate the main ideas of the first order formalism using a
toy example of the noninteracting neutral massive spin-zero
particles instead of the nonet of spin-1 resonances. The former
case shares all the main qualitative features with the latter and
due to its relative simplicity it suits well for our illustrative
purpose.

Spin-zero particles are usually described in terms of the real
massive scalar field with the Lagrangian
\begin{equation}  \label{Lfi}
\mathcal{L}_\phi =\frac 12\left( \partial \phi \right) ^2-\frac 12M^2\phi ^2\,.
\end{equation}
At the classical level, the corresponding Klein-Gordon equation is equivalent
to a pair of first order equations for scalar and vector fields $\phi$ and
$V_\mu$
\begin{equation}
\partial \phi -MV=0,\qquad \partial \cdot V+M\phi =0\,,  \label{equationKG}
\end{equation}
which can be derived from the ``mixed'' first order Lagrangian
\begin{equation}\label{LphiV}
\mathcal{L}_{\phi V}=MV\cdot \partial \phi -\frac 12M^2\phi ^2-\frac
12M^2V^2=MV\cdot \partial \phi -\mathcal{H}_{\phi V}\,.
\end{equation}
Here $\mathcal{H}_{\phi V}$ is the formal Legendre transformation
of the Lagrangian $\mathcal{L}_\phi$ with respect to the
derivatives $\partial _\mu \phi $ expressed in terms of its
``canonically adjoint variables'' $V_\mu M=\partial
\mathcal{L}_\phi /\partial \partial ^\mu \phi =\partial _\mu
\phi$.

On the other hand, eliminating $\phi $ from (\ref{equationKG}) we get
\begin{equation}
\partial _\mu \partial \cdot V+M^2V_\mu =0\,,
\end{equation}
which can be derived from the Lagrangian
\begin{equation}\label{LV}
\mathcal{L}_V=\frac 12\left( \partial \cdot V\right) ^2-\frac 12M^2V^2\,.
\end{equation}
This corresponds to an alternative and somewhat unusual possibility of
description of the free neutral massive spin-zero particles. The Lagrangian
$\mathcal{L}_{\phi V}$ can be in a sense interpreted as an interpolation
between the two possibilities represented by the Lagrangians $\mathcal{L}_\phi$
and $\mathcal{L}_V$. At the quantum level $\mathcal{L}_V$ leads to the
(covariant) propagator
\begin{equation}  \label{deltaV}
\mathrm{i}\Delta _F^V(p)_{\mu \nu }=\frac{\mathrm{i}}{M^2}\left( \frac{p_\mu
p_\nu }{p^2-M^2}-g_{\mu \nu }\right)\,,
\end{equation}
which differs from the propagator of the field $\partial _\mu \phi/M $ derived
from $\mathcal{L}_\phi$ by the presence of the contact term
$\text{$\sim$}g_{\mu\nu}$. This means that including the interaction the
descriptions based on $\mathcal{L}_\phi$ and $\mathcal{L}_V$ are not generally
equivalent, unless additional contact terms are introduced. This situation is
quite analogous to the case of Proca and tensor formalisms for spin-one
particles \cite{nase}.

The Lagrangian $\mathcal{L}_{\phi V}$ gives, besides the standard scalar
propagator and $\Delta_F^V(p)_{\mu\nu}$ propagator also the ``mixed'' one
\begin{equation}
\mathrm{i}\Delta _F^{\phi V}(p)_\mu = \mathrm{i}\Delta _F^{V\phi }(-p)_\mu =
\frac{\mathrm{i}}{p^2-M^2}\frac{\mathrm{i}p_\mu }M\,,
\end{equation}
which is a novel feature of this formalism.

Let us now add interaction with external sources (which mimic the chiral
building blocks here), \emph{i.e.} to (\ref{Lfi}), (\ref{LV}) and (\ref{LphiV}) we add
\begin{equation}
\mathcal{L}_\phi^{\text{int}}=j\phi -\frac 1MJ\cdot \partial \phi;\quad
\mathcal{L}_V^{\text{int}}=-\frac 1Mj\partial \cdot V+J\cdot V;\quad
\mathcal{L}_{\phi V}^{\text{int}}=j\phi +J\cdot V
\end{equation}
and suppose $j=O(p^2)$, $J=O(p^3)$. Integrating out the fields $\phi$ and/or
$V$ up to the order $O(p^6)$ we find out that in the case of first order
formalism
\begin{equation}
\mathcal{L}_{\phi V}^{\text{eff}} =\frac 1{2M^{2\,}}j^2+\frac 1{2M^4}\left(
\partial j\right) ^2+\frac 1{M^3}J\cdot \partial j+\frac 1{2M^{2\,}}J^2.
\end{equation}
The scalar field formalism misses the $O(p^6)$ contribution $\frac
1{2M^{2\,}}J^2$, while the vector field formalism includes this term,
however it does not yield any contribution of the order $O(p^4)$. The
situation here is just analogous to the little bit more complicated case of
the spin-1 resonances (cf. \cite{nase}).

\section{Example: vector formfactor}

\label{Secex}

Let us demonstrate the first order formalism for vector resonances on the
example of vector formfactor defined as:
\[
\langle \phi^b(p_1) \phi^c(p_2) | V^{\mu,a} | 0 \rangle = \mathcal{F}
(q^2)f^{abc}(p_2-p_1)^\mu\,.
\]
Here $\phi$s represent the Goldstone bosons and $q^\mu = (p_1+p_2)^\mu$. Using
the concrete form of the interaction Lagrangian and the Feynman rules given in
\cite{nase} we get at the tree level\footnote{Note that, formally, for
$f_V=g_V=0$ we recover the result of the tensor formalism, while for
$F_V=G_V=0$ the result of the Proca formalism is obtained}
\[
\mathcal{F}(q^2) = 1 - \frac{f_V g_V}{F^2}\frac{q^4}{q^2-M_V^2} - \frac{
F_VG_V}{F^2} \frac{q^2}{q^2-M_V^2} + \frac{F_Vg_V+f_VG_V}{M_V F^2}\frac{q^4}{
q^2-M_V^2}\,,
\]
where the independent terms are in the one-to-one correspondence with Fig.\ref{fig1}.
\newsavebox{\prop} \savebox{\prop}{\raisebox{-3pt}[0pt]{{
\epsfig{file=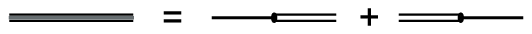}}}}
\begin{figure}[htb]
\begin{center}
\epsfig{file=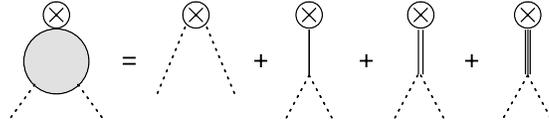}
\end{center}
\caption{Diagram representation of the vector formfactor. Single line stands
for Proca propagator, double line represents tensor propagator and dash
lines correspond to Goldstone bosons. Filled line is the combination of
mixed propagators: \usebox{\prop} as defined in \protect\cite{nase}.}
\label{fig1}
\end{figure}

High energy conditions requires that the formfactor vanishes for
$q^2\rightarrow \infty$. This implies the following conditions at the leading
order in $1/N_C$:

\begin{equation}  \label{dvepodm}
F_VG_V=F^2\,, \qquad g_V F_V (f_V M_V - F_V) = F^2 f_V\,.
\end{equation}
In the same way, imposing the short-distance constraints of $VVP$ correlator we
get $f_V = 0$ (for details see \cite{nase}) and the last condition of
(\ref{dvepodm}) simplifies to $g_V =0$.

\section{Conclusion}

In this paper, we have briefly  discussed  various ways of the
field-theoreti\-cal descriptions of the spin-1 resonances. Besides the most
common Proca and antisymmetric tensor field formalisms we have also assumed the
new first order formalism introduced recently in \cite{nase}. Typical features
of the latter have been illustrated using a toy example of the spin-0
particles. This simplification shares all the  qualitative features with the
spin-1 case: (\emph{i})~description using a pair of fields with different
Lorentz transformation properties, (\emph{ii})~Feynman rules with mixed
propagators and (\emph{iii})~richer structure of the effective action obtained
by means of integrating out the heavy fields. It might be interesting to
further investigate the relevance of this formalism for the description of
spin-0 resonances within $\chi RT$.

As an illustration of the first order formalism for spin-1 resonances we have
explicitly calculated the vector formfactor of pion and discussed the
consistency of the result with high energy behaviour. This supplies us with
simple conditions without necessity of adding further contact terms. Moreover
using the short-distance constraints of $VVP$ correlator (cf. \cite {nase}),
the same result as in tensor formalism is obtained.

\end{document}